\documentstyle[newarcrc,fleqn]{article}
\input{psfig.sty}

\title{New evolutionary scenarios for short orbital period CVs.}

\author{Retter A., Naylor T. \address{Department of Astronomy, Keele 
University, Staffs., ST5 5BG, UK; \\ ar@astro.keele.ac.uk; 
timn@astro.keele.ac.uk}}

\begin{document}
% typeset front matter
\maketitle

\begin{abstract}

We suggest new evolutionary scenarios for non-magnetic short orbital period 
CVs. The first model is the analogy of the `hibernation scenario' or the 
`modern hibernation scenario'. The second one is an extension of Mukai 
\& Naylor (1995) ideas. All models imply a tight connection between 
permanent superhump systems and classical novae. We highlight the 
significance of the observed evolution of V1974 Cyg, which might pose a 
major problem to Mukai \& Naylor concept.

\end{abstract}

\section{Introduction}

The `hibernation scenario' (Shara 1989 for review) suggests that dwarf 
novae $\rightarrow$ nova-likes $\rightarrow$ novae $\rightarrow$ nova-likes 
$\rightarrow$ dwarf novae $\rightarrow$ hibernation $\rightarrow$ dwarf 
novae etc. However, it was later proposed that the hibernation stage 
($\dot{M} \sim 0$) might not exist at all (Livio 1989), thus dwarf novae 
$\rightarrow$ nova-likes $\rightarrow$ novae $\rightarrow$ nova-likes 
$\rightarrow$ dwarf novae... The typical time scales for the transitions 
were estimated as a few centuries. 

An alternative view to the `hibernation scenario' was presented by Mukai and 
Naylor (1995). They suggested that nova-likes and dwarf novae constitute
different classes of pre-nova systems. Therefore, there are two 
possibilities:
1. nova-likes $\rightarrow$ novae $\rightarrow$ nova-likes...
2. dwarf novae $\rightarrow$ novae $\rightarrow$ dwarf novae...
Nova-likes should have more frequent nova outbursts than dwarf novae because 
their mass transfer rates are higher. The critical mass for the 
thermonuclear runaway is thus achieved much faster. According to this model,
a long-term evolution between the two phases might occur.

\section{Discussion}

\subsection{Permanent superhump novae}

We note the similarity between long and short orbital period CVs. Permanent 
superhump systems have thermally stable accretion discs as do nova-likes, 
while SU UMa and U Gem systems are thermally unstable.

So far only two non-magnetic novae below the period gap have been discovered - 
CP Pup 1942 and V1974 Cyg 1992. Both have permanent superhumps in their 
light curves. To these systems, we naturally add V603 Aql 1918, the third 
permanent superhump nova, whose binary period is just the other side of 
the gap (Retter \& Leibowitz 1998). These three objects show clearly that 
certain classical novae become permanent superhump systems. Since the three 
post-novae are permanent superhumpers, their accretion discs should be 
thermally stable. When we compare, however, the pre-outburst luminosities 
with the post-nova values, various types of behaviour are discovered. V603 
Aql seems to have returned exactly to its pre-outburst magnitude. The upper 
limit on the brightness of the progenitor of CP Pup (Warner 1995) shows 
that it was fainter than its post outburst quiescent value, but prevents 
a precise decision concerning the thermal stability of the pre-nova. 
V1974 Cyg is the most interesting case among the three. Retter \& Leibowitz 
(1998) argued that the pre-nova was faint, and therefore should have been a 
dwarf nova (SU UMa system) with a thermally unstable accretion disc. It is 
thus the only clear case of a classical nova that has changed its thermal 
stability state -- a change to the thermally stable state from the thermally 
unstable state have been caused by the nova outburst. CP Pup might be a second
example of this transition.

\subsection{Summary of relevant observations}

Observations of novae have shown the following:\\
1. Most systems probably have only a short cycle (nova-likes $\rightarrow$ 
novae $\rightarrow$ nova-likes... -- i.e. Robinson 1975).
Two selection effects might, however, be involved -- brighter novae are 
better covered, and a longer observational base line might show a larger 
cycle.\\
2. There are two clear examples of novae that have turned into dwarf novae -- 
V446 Her (Honeycutt et al. 1998) and GK Per (Sabbadin \& Bianchini 1983).
GK Per is, however, not a typical nova.\\
3. There is at least one example of a nova like (permanent superhump system) 
-- V1974 Cyg, that should have been a dwarf nova (SU UMa system) before 
its nova event (Retter \& Leibowitz 1998).

\subsection{New evolutionary scenarios}

We suggest that most (or perhaps all) non-magnetic short orbital period 
novae should evolve into permanent superhump systems. We also propose that 
most (or all) permanent superhump systems are ex-novae. An indirect 
evidence, supporting the last claim, is the fact that permanent superhumps 
have been detected in a few SW Sex stars (Patterson 1999), and three SW Sex 
candidates are actually old novae (Hoard 1998). Further evidence for this 
idea comes for the possible identification of BK Lyn, a permanent superhump 
system (Patterson 1999), with a Chinese guest star, which erupted in 101 
(Hertzog 1986).  

We further suggest that evolutionary scenarios, similar to those offered for
the long orbital period CVs, are applicable to the short orbital period 
systems, as well. The equivalence of the `hibernation scenario' is:
SU UMa systems ($\rightarrow$ permanent superhump systems) $\rightarrow$ 
novae $\rightarrow$ permanent superhumpers $\rightarrow$ SU UMa systems
$\rightarrow$ hibernation $\rightarrow$ SU UMa systems... The analogy of 
the `modern hibernation scenario' is: SU UMa systems ($\rightarrow$ 
permanent superhump systems) $\rightarrow$ novae $\rightarrow$ permanent 
superhumpers $\rightarrow$ SU UMa systems... The extension of the two options
of Mukai and Naylor (1995) ideas is: permanent superhumpers $\rightarrow$ 
novae $\rightarrow$ permanent superhumpers... and SU UMa systems $\rightarrow$
 novae $\rightarrow$ SU UMa systems... The observed evolution of V1974 Cyg 
is consistent with this view only if the CV was caught in a very specific
point in its long term evolution -- a transition from the faint stage to 
the bright stage. 

% evolutionary scenario equivalent to the different view of Mukai 
%\& Naylor (1995). We can, however, meditate a third combined scenario, in 
%which CVs spend a part of their life time in a high state, as permanent 
%superhump systems erupting into novae and backwards, but during other parts 
%of their course they erupt to novae from the SU UMa state, and decay to 
%become dwarf novae again with a possible additional hibernation phase. 
%Similar scenario is proposed for CVs with orbital periods above the period 
%gap.

\end{document}